\documentclass[reprint,nofootinbib,...]{revtex4-1} 
\usepackage{amsmath,amssymb,amsfonts,bm,graphicx,epsf,colordvi,bbm,pifont,siunitx,enumitem,xcolor,multirow}
\usepackage[caption=false]{subfig}
\pdfoutput=1
\usepackage[hyperfootnotes=false]{hyperref}
\hypersetup{
 colorlinks=true,
 citecolor=blue,
 linkcolor=blue,
 urlcolor=blue}
 \usepackage[normalem]{ulem}

\usepackage{amsmath}%
\usepackage{amsthm,amssymb}
\usepackage{graphicx}
\usepackage{epstopdf}
\usepackage{xcolor}
\usepackage{algcompatible}
\usepackage[size=small]{caption}
\usepackage{etoolbox}
\usepackage{booktabs}
\usepackage{multirow}
\usepackage[utf8]{inputenc}
\usepackage[colorinlistoftodos]{todonotes}
\AtBeginEnvironment{algorithm}{\noindent\hrulefill\par\nobreak\vskip-5pt}
\usepackage{newfloat}

\DeclareFloatingEnvironment[
    fileext=loa,
    listname=List of Algorithms,
    name=ALGORITHM,
    placement=tbhp,
]{algorithm}
\DeclareCaptionFormat{algorithms}{\vskip-15pt\hrulefill\par#1#2#3\vskip-6pt\hrulefill}
\algrenewcommand{\algorithmiccomment}[1]{\hskip3em// {\it #1}}

\usepackage[braket]{qcircuit}
\usepackage{tikz}
\usetikzlibrary{arrows,shapes}
\usetikzlibrary{trees}
\usetikzlibrary{matrix,arrows} 	
\tikzstyle{block} = [draw, rectangle, 
    minimum height=3em, minimum width=6em]

\usepackage{epsfig}
\usepackage{comment}

\graphicspath{{figures/}}

\begin{document} 
\title{A Quantum Algorithm to Efficiently Sample from Interfering Binary Trees} 

\author{Davide Provasoli}
\email{davideprovasoli@lbl.gov}

\affiliation{Physics Division, Lawrence Berkeley National Laboratory, Berkeley, CA 94720, USA}

\author{Benjamin Nachman}
\email{bpnachman@lbl.gov}

\affiliation{Physics Division, Lawrence Berkeley National Laboratory, Berkeley, CA 94720, USA}

\author{Wibe A. de Jong}
\email{WAdeJong@lbl.gov}

\affiliation{Computational Research Division, Lawrence Berkeley National Laboratory, Berkeley, CA 94720, USA}

\author{Christian Bauer}
\email{cwbauer@lbl.gov}

\affiliation{Physics Division, Lawrence Berkeley National Laboratory, Berkeley, CA 94720, USA}

\newcommand{\ch}[1]{{\color{magenta} #1}}
\newcommand{\bn}[1]{{\color{blue} #1}}

\begin{abstract}
Quantum computers provide an opportunity to efficiently sample from probability distributions that include non-trivial interference effects between amplitudes. Using a simple process wherein all possible state histories can be specified by a binary tree, we construct an explicit quantum algorithm that runs in polynomial time to sample from the process once.  The corresponding naive Markov Chain algorithm does not produce the correct probability distribution and an explicit classical calculation of the full distribution requires exponentially many operations.  However, the problem can be reduced to a system of two qubits with repeated measurements, shedding light on a quantum-inspired efficient classical algorithm. 
\end{abstract}

\maketitle

\section{Introduction}
\label{sec:intro}

Quantum algorithms are promising for various industrial and scientific applications because of their capacity to explore exponentially many states with a polynomial number of quantum bits.  One of the most well-studied classes of quantum algorithms is the quantum walk~\cite{PhysRevA.48.1687}.  Like the classical random walk, the quantum variants have found widespread use for enhancing a variety of quantum calculations and simulations~\cite{VenegasAndraca:2012fh,review}.  While quantum walks are fundamentally different from classical random walks, there are limits in which the quantum algorithm approaches the classical one~\cite{PhysRevA.93.062316}.

A useful feature of a classical random walk is that it can be efficiently simulated using a Markov Chain Monte Carlo (MCMC) because subsequent motion depends only on the current position and not the prior history.  This MC property is at the core of some algorithms that simulate many-body physical systems where the generative process is approximately local.  For such physical systems that also have important quantum properties, the speed from the MCMC is traded off against the accuracy of an inherently quantum simulation.  One such physical system is the parton shower in high energy physics~\cite{Patrignani:2016xqp}, where a quark or a gluon radiates a shower of nearly collinear quarks and gluons.  Genuine quantum effects can be approximated as corrections to the MCMC~\cite{Nagy:2014mqa}, but cannot be directly implemented efficiently in a classical MCMC approach.


Consider the following \textit{quantum tree}: at every step, a spin 1/2 particle can move one unit left or one unit right.  After $N$ steps, this system forms a binary tree with $2^N$ paths.  In contrast to a traditional quantum walk, we assume that the path is observable, so moving left and then right is not the same as moving right and then left.  For this reason, there is a 1-1 correspondence between the leaves of the tree and the path taken, and the space of measurement outcomes is more naturally $\{L,R\}^N$ than $\mathbb{Z}$.  

When the quantum amplitude for moving left is independent of the spin or if the spin changes deterministically with time, this tree can be efficiently and accurately simulated with a classical MCMC.  However, when either of these conditions are violated, a naive classical MCMC fails to produce the correct probability distribution over final states.  While quantum walks with time/space dependence have been studied in the literature~\cite{0305-4470-37-30-013,PhysRevA.73.062304,PhysRevLett.114.140502,PhysRevA.93.062316,1367-2630-20-8-083028} and there are some similarities to quantum algorithms for decision trees~\cite{PhysRevA.58.915}, our quantum tree requires a new approach.

In order to efficiently sample from the quantum tree, we introduce a new quantum algorithm that achieves an exponential speedup over an efficient classical calculation of the full final state probability distribution.  In addition, we provide an explicit quantum circuit which implements the algorithm and demonstrate its performance on a quantum computer simulation.  Interestingly, an equivalent quantum circuit involving only two qubits can be obtained if we use repeated measurements, and this shed light on a quantum inspired classical algorithm that is indeed an efficient MCMC.

This paper is organized as follows.  Section~\ref{sec:classical} introduces the quantum tree and illustrates how naive classical algorithms cannot efficiently sample from its probability distribution.  A solution to this problem is introduced in Sec.~\ref{sec:quantum} using a quantum algorithm.  An explicit implementation of the quantum circuit is described in Sec.~\ref{sec:implementation} and numerical results are presented in Sec.~\ref{sec:results}.  An efficient quantum-inspired classical MCMC is introduced in Sec.~\ref{sec:quantumclassical}.  The paper ends with conclusions and future outlook in Sec.~\ref{sec:conclusions}.

\section{A Classical Challenge}
\label{sec:classical}
Consider a tree like the one shown in Fig.~\ref{fig:tree}, where the quantum amplitude of a node $n$ is given by $A_L(n)$ when going left and $A_R(n)$ when going right.  The amplitude for reaching a given leaf is the product over the nodes from its history $\lambda \in\{L,R\}^N$: $A_\text{leaf}=\prod_{n=1}^N A_{\lambda_n}(n)$.  The probability of paths through the tree (uniquely specified by a leaf) are distributed according to $\Pr(\text{path})\propto |A_\text{leaf}|^2$.  One can efficiency sample from this distribution in linear time classically using a MCMC algorithm: at each step, move left or right with a probability given by $|A_{L/R}(n)|^2$.  

Now, consider the following change to the tree: there is a spin state associated with each depth.  Only the spin at the leaf is observable and the amplitudes $A_L$ and $A_R$ depend on the state of the spin.  Now, there are many possible paths that correspond to reaching a single leaf.  One way to visualize this is illustrated in Fig.~\ref{fig:tree2}.   There are two copies of the tree, one for spin up and one for spin down.  At each step, the system can move between trees or stay on the same tree and then move left or right.  The observable final state is the leaf location and the final tree (spin).  The amplitudes for going left and right are now spin-dependent.  At a given step, the eight possible amplitudes are $A_{h}^{s_1,s_2}(n)$ for $h\in\{L,R\}$ and $s_i\in\{\uparrow,\downarrow\}$, where $s_1$ is the initial spin and $s_2$ is the final spin. Since only the final spin is observable, the amplitude to transition from spin $s_0$ to $s_N$ is given by
\begin{align}
\label{eq:amp_combination}
A_{s_0, s_N} = \sum_{\stackrel{\vec{s}' \in \{\downarrow,\uparrow\}^N}{s'_0 = s_0, s'_N = s_N}} \prod_{n=1}^{N} A_{\lambda_n}^{s'_{n-1}, s'_n}(n)
\,.
\end{align}


While there may be multiple applications of this quantum tree, one motivation is the parton shower in quantum chromodynamics (QCD) where quarks or gluons radiate gluons (going left in the tree) at decreasing angles (deeper $n$).  The connection with QCD is not exact but the work presented here is a step toward an inherently quantum parton shower algorithm.  

\begin{figure}[h!]
\begin{tikzpicture}[line width=1.5 pt]
\begin{scope}[shift={(-1.3,0)},rotate=-90]
\draw (0,0) -- (1,1);
\draw (0,0) -- (1,-1);
\node at (0.4,1.1) {\color{black}$A_{R}$};
\node at (0.4,-1.1) {\color{black}$A_{L}$};

\draw (1,1) -- (1+0.5,1+0.5);
\draw (1,1) -- (1+0.5,1-0.5);

\draw (1,-1) -- (1+0.5,-1+0.5);
\draw (1,-1) -- (1+0.5,-1-0.5);

\draw (1+0.5,1+0.5) -- (1+0.5+0.25,1+0.5+0.25);
\draw (1+0.5,1+0.5) -- (1+0.5+0.25,1+0.5-0.25);

\draw (1+0.5,1-0.5) -- (1+0.5+0.25,1-0.5+0.25);
\draw (1+0.5,1-0.5) -- (1+0.5+0.25,1-0.5-0.25);

\draw (1+0.5,-1+0.5) -- (1+0.5+0.25,-1+0.5+0.25);
\draw (1+0.5,-1+0.5) -- (1+0.5+0.25,-1+0.5-0.25);

\draw (1+0.5,-1-0.5) -- (1+0.5+0.25,-1-0.5+0.25);
\draw (1+0.5,-1-0.5) -- (1+0.5+0.25,-1-0.5-0.25);
\end{scope}
 \end{tikzpicture}
 \caption{The rightmost nodes of the above binary tree (leaves) uniquely correspond to trajectories in $\{L,R\}^N$ where $L$ represents going left and $R$ represents going right at a given node.  As a generative model, trajectories are sampled according to the square of the quantum amplitude of the path through the tree.}
 \label{fig:tree}
 \end{figure}
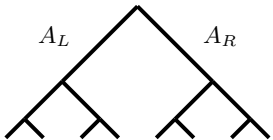

\begin{figure}[h!]
\begin{tikzpicture}[line width=1.5 pt]
\node at (-2.2+4.3,-0.6+0.2) {\color{red}$A_{L}^{\uparrow\uparrow}$};
\node at (-0.4+4.3,-0.6+0.2) {\color{red}$A_{R}^{\uparrow\uparrow}$};
\node at (-2.2,-0.6+0.2) {$A_{L}^{\downarrow\downarrow}$};
\node at (-0.4,-0.6+0.2) {$A_{R}^{\downarrow\downarrow}$};
\node at (-1.3,-1.1) {\color{red}$A_{L}^{\uparrow\downarrow}$};
\node at (0.4,-1.1) {\color{red}$A_{R}^{\uparrow\downarrow}$};
\node at (1.3,-1.1) {\color{black}$A_{L}^{\downarrow\uparrow}$};
\node at (-1.3+4.3,-1.1) {\color{black}$A_{R}^{\downarrow\uparrow}$};
\begin{scope}[shift={(-1.3,0)},rotate=-90]
\draw (0,0) -- (1,1);
\draw (0,0) -- (1,-1);

\draw[black!50] (1,1) -- (1+0.5,1+0.5);
\draw[black!50] (1,1) -- (1+0.5,1-0.5);

\draw[black!50] (1,-1) -- (1+0.5,-1+0.5);
\draw[black!50] (1,-1) -- (1+0.5,-1-0.5);

\draw[black!50] (1+0.5,1+0.5) -- (1+0.5+0.25,1+0.5+0.25);
\draw[black!50] (1+0.5,1+0.5) -- (1+0.5+0.25,1+0.5-0.25);

\draw[black!50] (1+0.5,1-0.5) -- (1+0.5+0.25,1-0.5+0.25);
\draw[black!50] (1+0.5,1-0.5) -- (1+0.5+0.25,1-0.5-0.25);

\draw[black!50] (1+0.5,-1+0.5) -- (1+0.5+0.25,-1+0.5+0.25);
\draw[black!50] (1+0.5,-1+0.5) -- (1+0.5+0.25,-1+0.5-0.25);

\draw[black!50] (1+0.5,-1-0.5) -- (1+0.5+0.25,-1-0.5+0.25);
\draw[black!50] (1+0.5,-1-0.5) -- (1+0.5+0.25,-1-0.5-0.25);

\draw[dashed,black!30,line width=0.25mm] (0,0) -- (1,3.3);
\draw[dashed,black!30,line width=0.25mm] (0,0) -- (1,5.3);

\draw[dashed,red!30,line width=0.25mm] (0,4.3) -- (1,1);
\draw[dashed,red!30,line width=0.25mm] (0,4.3) -- (1,-1);

\end{scope}
\begin{scope}[shift={(3,0)},rotate=-90]
\draw [red] (0,0) -- (1,1);
\draw [red] (0,0) -- (1,-1);

\draw[red!50] (1,1) -- (1+0.5,1+0.5);
\draw [red!50](1,1) -- (1+0.5,1-0.5);

\draw [red!50](1,-1) -- (1+0.5,-1+0.5);
\draw[red!50](1,-1) -- (1+0.5,-1-0.5);

\draw[red!50] (1+0.5,1+0.5) -- (1+0.5+0.25,1+0.5+0.25);
\draw[red!50] (1+0.5,1+0.5) -- (1+0.5+0.25,1+0.5-0.25);

\draw[red!50] (1+0.5,1-0.5) -- (1+0.5+0.25,1-0.5+0.25);
\draw[red!50] (1+0.5,1-0.5) -- (1+0.5+0.25,1-0.5-0.25);

\draw[red!50] (1+0.5,-1+0.5) -- (1+0.5+0.25,-1+0.5+0.25);
\draw [red!50](1+0.5,-1+0.5) -- (1+0.5+0.25,-1+0.5-0.25);

\draw[red!50] (1+0.5,-1-0.5) -- (1+0.5+0.25,-1-0.5+0.25);
\draw[red!50] (1+0.5,-1-0.5) -- (1+0.5+0.25,-1-0.5-0.25);
\end{scope}

 \end{tikzpicture}
  \caption{The same setup as in Fig.~\ref{fig:tree}, except that now there is a spin state associated with every depth in the tree.  This can be represented by two trees: one for spin down (left) and one for spin up (right).  The system can move between trees, but only the final tree (spin) and leaf are observable. The eight possible amplitudes for a given step are indicated with $A_{L/R}^{s_1,s_2}$, where $s_1$ is the initial spin and $s_2$ is the final spin.}
 \label{fig:tree2}
 \end{figure}
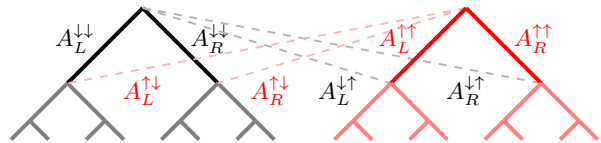

The quantum tree including the full interference effects caused by cross-terms in the sum over all spin histories for a given leaf cannot be implemented in a naive MCMC that extends the one from Fig.~\ref{fig:tree} where each possibility was sampled at each step.  One method for correctly sampling from the distribution of leafs and final spins is to sum over all paths to compute the probabilities for each state.  For a tree of depth $N$, the calculation of the total amplitude would naively scale as $4^N$ since there are 4 possibilities at every node: move left and flip the spin, move right and flip the spin, move left and do not flip the spin, move right and do not flip the spin. 


One way to efficiently calculate the probability distribution is to represent the problem as a set of matrix multiplications.  To see this, consider the leaf corresponding to never taking the left branch.  The probability for the two possible states (spin up or spin down) requires summing over all possible spin trajectories.  If the initial spin is $\ket{i}$ for $\ket{\downarrow}=\begin{pmatrix}1 \cr 0\end{pmatrix}$ and $\ket{\uparrow}=\begin{pmatrix}0 \cr 1\end{pmatrix}$, then one can compute the full probability distribution of the final spin $\ket{f}$ by matrix multiplication:

\begin{align}
\ket{f} =\prod_{n=1}^N \Delta(n) \ket{i}, \Delta(n)= \left( \begin{array}{cc} A_{R}^{\downarrow\downarrow}(n)  &A_{R}^{\uparrow\downarrow}(n) \\ A_{R}^{\downarrow\uparrow}(n)& A_{R}^{\uparrow\uparrow}(n)\end{array}\right)
\,.\end{align}

\noindent Therefore, the amplitude for the right-move only case can be computed with $\mathcal{O}(N)$ multiplications.   The same logic applies to the calculation of the amplitude for exactly one left branch at step $k$:

\begin{align}
\label{sec:onemission}
\ket{f} = \prod_{n=k+1}^N \Delta(n) \times A(k) \times\prod_{n=1}^{k-1}\Delta(n)\ket{i}
\,,\end{align}

\noindent where

\begin{align}
A(k)= \left( \begin{array}{cc} A_{L}^{\downarrow\downarrow}(k)  & A_{L}^{\uparrow\downarrow}(k) \\ A_{L}^{\downarrow\uparrow}(k) &A_{L}^{\uparrow\uparrow}(k) \end{array}\right)
\,.
\end{align}

\noindent Equation~\ref{sec:onemission} is also inefficient when considering all $1\leq k\leq N$, because many products can be reused from one $k$ to another.  However, even with the maximal amount of reuse, there must be at least one matrix multiplication per $k$ value.  By the same logic, there must be at least one matrix multiplication for every fixed number of left branchings.  There are a total of $2^N$ leaves and therefore the minimum number of matrix multiplications scales exponentially with $N$.  Particular re-use schemes can be deployed to show that the scaling is $2^N$ and to calculate the coefficient of the exponential scaling. In the next section will show that there exists a quantum algorithm that can distribute events from this probability distribution, where a single event can be generated in polynomial time. This therefore provides an exponential speedup over the naive classical approach.

\section{A Quantum Solution}
\label{sec:quantum}
\subsection{Rotating to a new basis}
Let us write the evolution in Eq.~\eqref{eq:amp_combination} in terms of the following two unitary transformations, pertaining to one step starting on the $\ket{\downarrow}$ tree and one step starting on the $\ket{\uparrow}$ tree:
\begin{align}
\label{eq:step}
\ket{\downarrow} \rightarrow A_{L}^{\downarrow\downarrow}\ket{L}\ket{\downarrow}+A_{L}^{\downarrow\uparrow}\ket{L}\ket{\uparrow}+A_{R}^{\downarrow\downarrow}\ket{R}\ket{\downarrow}+A_{R}^{\downarrow\uparrow}\ket{R}\ket{\uparrow} \nonumber \\
\ket{\uparrow} \rightarrow A_{L}^{\uparrow\uparrow}\ket{L}\ket{\uparrow}+A_{L}^{\uparrow\downarrow}\ket{L}\ket{\downarrow}+A_{R}^{\uparrow\uparrow}\ket{R}\ket{\uparrow}+A_{R}^{\uparrow\downarrow}\ket{R}\ket{\downarrow}
\,,
\end{align}
where the amplitudes must satisfy the unitarity conditions:
\begin{align}
\label{eq:unitarity}
{A_{L}^{\downarrow\downarrow}}^2+{A_{L}^{\downarrow\uparrow}}^2+{A_{R}^{\downarrow\downarrow}}^2+{A_{R}^{\downarrow\uparrow}}^2 = 1 \\ \nonumber
{A_{L}^{\uparrow\uparrow}}^2+{A_{L}^{\uparrow\downarrow}}^2+{A_{R}^{\uparrow\uparrow}}^2+{A_{R}^{\uparrow\downarrow}}^2 = 1
\,.
\end{align}
This evolution will produce interference terms, since we can reach the same state in more than one way, and as previously mentioned it cannot be implemented with a simple naive MCMC. 
We would like to rotate to a new basis
\begin{align}
\label{eq:basis_rotation}
\ket{\downarrow'} = \cos{\lambda}\ket{\downarrow}-\sin{\lambda}\ket{\uparrow} \nonumber \\
\ket{\uparrow'} = \sin{\lambda}\ket{\downarrow}+\cos{\lambda}\ket{\uparrow}
\,,
\end{align}
such that one evolution step looks like
\begin{align}
\label{eq:new_step}
\ket{\downarrow'}\rightarrow \widetilde{A}_{L}^{\downarrow\downarrow}\ket{L}\ket{\downarrow'}+\widetilde{A}_{R}^{\downarrow\downarrow}\ket{R}\ket{\downarrow'} \nonumber \\
\ket{\uparrow'}\rightarrow \widetilde{A}_{L}^{\uparrow\uparrow}\ket{L}\ket{\uparrow'}+\widetilde{A}_{R}^{\uparrow\uparrow}\ket{R}\ket{\uparrow'}
\,,
\end{align}
with unitarity conditions
\begin{align}
\label{eq:unitarity2}
(\widetilde{A}_{L}^{\downarrow\downarrow})^2 + (\widetilde{A}_{R}^{\downarrow\downarrow})^2 = 1 \nonumber \\
(\widetilde{A}_{L}^{\uparrow\uparrow})^2 + (\widetilde{A}_{R}^{\uparrow\uparrow})^2 = 1
\,.
\end{align}
In the new basis the two trees decouple and the evolution becomes simple, meaning our quantum states evolve at each step by going either right or left, but they can no longer go in between trees.  The original system had six degrees of freedom (8 amplitudes and two unitary conditions given in Eq.~\eqref{eq:unitarity}) while the new system has only three degrees of freedom (one from $\lambda$, four amplitudes and two unitary conditions given in Eq.~\eqref{eq:unitarity2}).  This means that this basis switch is only possible for a subset of cases for the original problem.  Since these cases admit a simple quantum algorithm, we focus on these and leave the general case for future studies.

In order to find the correct rotation angle $\lambda$ to implement Eq.~\eqref{eq:basis_rotation}, we must solve:
\begin{align}
\label{eq:solve_down}
\ket{\downarrow'} \rightarrow & \cos{\lambda} ( A_{L}^{\downarrow\downarrow}\ket{L}\ket{\downarrow}+A_{L}^{\downarrow\uparrow}\ket{L}\ket{\uparrow}+A_{R}^{\downarrow\downarrow}\ket{R}\ket{\downarrow} \nonumber \\ 
& +A_{R}^{\downarrow\uparrow}\ket{R}\ket{\uparrow}) - \sin{\lambda} (A_{L}^{\uparrow\uparrow}\ket{L}\ket{\uparrow}+A_{L}^{\uparrow\downarrow}\ket{L}\ket{\downarrow} \nonumber \\
& +A_{R}^{\uparrow\uparrow}\ket{R}\ket{\uparrow}+A_{R}^{\uparrow\downarrow}\ket{R}\ket{\downarrow}) \nonumber \\
& = \ket{L}\Big[\cos{\lambda}A_{L}^{\downarrow\downarrow}\ket{\downarrow}+\cos{\lambda}A_{L}^{\downarrow\uparrow}\ket{\uparrow} - \sin{\lambda}A_{L}^{\uparrow\uparrow}\ket{\uparrow} \nonumber \\
& - \sin{\lambda}A_{L}^{\uparrow\downarrow}\ket{\downarrow}\Big] + \ket{R}\Big[\cos{\lambda}A_{R}^{\downarrow\downarrow}\ket{\downarrow} \nonumber \\
& + \cos{\lambda}A_{R}^{\downarrow\uparrow}\ket{\uparrow} - \sin{\lambda}A_{R}^{\uparrow\uparrow}\ket{\uparrow} - \sin{\lambda}A_{R}^{\uparrow\downarrow}\ket{\downarrow}\Big]
\,.
\end{align}
Focusing on the term proportional to $\ket{L}$, from Eq.~\eqref{eq:solve_down} we have
\begin{align}
&  \widetilde{A}_{L}^{\downarrow\downarrow}\ket{\downarrow'} =  \widetilde{A}_{L}^{\downarrow\downarrow} \big(\cos{\lambda}\ket{\downarrow}- \sin{\lambda}\ket{\uparrow}\big)\nonumber \\
& = \cos{\lambda}A_{L}^{\downarrow\downarrow}\ket{\downarrow}+\cos{\lambda}A_{L}^{\downarrow\uparrow}\ket{\uparrow} - \sin{\lambda}A_{L}^{\uparrow\uparrow}\ket{\uparrow} \nonumber \\
& - \sin{\lambda}A_{L}^{\uparrow\downarrow}\ket{\downarrow}
\,,
\end{align}
so that we get the following two equations:
\begin{align}
\label{eq:left_down}
\cos{\lambda}\widetilde{A}_{L}^{\downarrow\downarrow} = \cos{\lambda}A_{L}^{\downarrow\downarrow}- \sin{\lambda}A_{L}^{\uparrow\downarrow} \nonumber \\
-\sin{\lambda} \widetilde{A}_{L}^{\downarrow\downarrow}  = \cos{\lambda}A_{L}^{\downarrow\uparrow} - \sin{\lambda}A_{L}^{\uparrow\uparrow}
\end{align}
If we multiply the top equation by $\sin{\lambda}$, the bottom equation by $\cos{\lambda}$ and add them we get
\begin{align}
\label{eq:eq1L}
\cos{\lambda}\sin{\lambda}(A_L^{\downarrow\downarrow}-A_L^{\uparrow\uparrow}) + \cos{\lambda}^2(A_L^{\uparrow\downarrow} + A_L^{\downarrow\uparrow}) - A_L^{\uparrow\downarrow} = 0
\,.
\end{align}
Now if we repeat the same process with the transformation of $\ket{\uparrow'}$ and once again we focus on the terms proportional to $\ket{L}$ we obtain
\begin{align}
\label{eq:left_up}
\sin{\lambda}\widetilde{A}_{L}^{\uparrow\uparrow} = \sin{\lambda}A_{L}^{\downarrow\downarrow} + \cos{\lambda}A_{L}^{\uparrow\downarrow} \nonumber \\
\cos{\lambda} \widetilde{A}_{L}^{\uparrow\uparrow}  = \sin{\lambda}A_{L}^{\downarrow\uparrow} + \cos{\lambda}A_{L}^{\uparrow\uparrow}
\end{align}
and
\begin{align}
\label{eq:eq2L}
\cos{\lambda}\sin{\lambda}(A_L^{\downarrow\downarrow}-A_L^{\uparrow\uparrow}) + \cos{\lambda}^2(A_L^{\uparrow\downarrow} + A_L^{\downarrow\uparrow}) - A_L^{\downarrow\uparrow} = 0
\,.
\end{align}
Then Eq. \eqref{eq:eq1L} and Eq. \eqref{eq:eq2L} imply
\begin{align}
\label{condition1}
A_L^{\downarrow\uparrow} = A_L^{\uparrow\downarrow} \equiv A_L
\end{align}
and they become
\begin{align}
\label{eq:eqL}
\cos{\lambda}\sin{\lambda}(A_L^{\downarrow\downarrow}-A_L^{\uparrow\uparrow}) + \cos{2\lambda}A_L= 0
\,.
\end{align}
We can now solve for $\lambda$ in terms of $A_L$, $A_L^{\uparrow\uparrow}$ and $A_L^{\downarrow\downarrow}$, which are  free parameters we will specify in the unrotated basis, and use the result to solve for  $\widetilde{A}_{L}^{\downarrow\downarrow}$ and $\widetilde{A}_{L}^{\uparrow\uparrow}$ in Eqs. \ref{eq:left_down} and \ref{eq:left_up} . When we do so we get
\begin{align}
\label{A_tildes}
\widetilde{A}_{L}^{\downarrow\downarrow} & =  A_L^{\downarrow\downarrow} \nonumber \\
& - \frac{\sqrt{4 A_L^2 (A_L^{\downarrow\downarrow}-A_L^{\uparrow\uparrow})^2 + (A_L^{\downarrow\downarrow}-A_L^{\uparrow\uparrow})^4 } + (A_L^{\downarrow\downarrow}-A_L^{\uparrow\uparrow})^2}{2 (A_L^{\downarrow\downarrow}-A_L^{\uparrow\uparrow})} \\
\widetilde{A}_{L}^{\uparrow\uparrow} & = A_L^{\downarrow\downarrow} \nonumber \\
& + \frac{2 A_L^2 (A_L^{\downarrow\downarrow}-A_L^{\uparrow\uparrow})}{\sqrt{4 A_L^2 (A_L^{\downarrow\downarrow}-A_L^{\uparrow\uparrow})^2 + (A_L^{\downarrow\downarrow}-A_L^{\uparrow\uparrow})^4 } + (A_L^{\downarrow\downarrow}-A_L^{\uparrow\uparrow})^2}
\,.
\end{align}
We can then find $\widetilde{A}_{R}^{\uparrow\uparrow}$ and $\widetilde{A}_{R}^{\downarrow\downarrow}$ from unitarity conditions in \ref{eq:unitarity2}. Of course we could have performed the same derivation focusing on terms proportional to $\ket{R}$ instead, in which case, instead of eq. \ref{eq:eqL}, we would have found 
\begin{align}
\label{condition2}
A_R^{\downarrow\uparrow} = A_R^{\uparrow\downarrow} \equiv A_R
\end{align}
and
\begin{align}
\label{eq:eqR}
\cos{\lambda}\sin{\lambda}(A_R^{\downarrow\downarrow}-A_R^{\uparrow\uparrow}) + \cos{2\lambda}A_R= 0
\,.
\end{align}


\subsection{Tree evolution as an efficient quantum algorithm}

We now introduce a quantum algorithm which can solve the system introduced in the previous section in polynomial time. The algorithm implements the change of basis discussed above, it evolves the system in the decoupled basis and then rotates back to the original basis, creating interferences between all the possible paths which lead to the same final leaf and spin.

To illustrate the algorithm consider a tree of the kind illustrated in Fig.~\ref{fig:tree2} with $N$ total nodes and a spin degree of freedom. The state which is evolved in our quantum circuit is given by
\begin{align}
\ket{\Psi_{n,N}} & = \ket{s}\ket{\lambda_1 \lambda_2 \ldots \lambda_n \ldots \lambda_N}
\nonumber\\ & \equiv  \ket{\psi_{n,N}}
\,,
\end{align}
where $n$ denotes how many steps have occured and the combination of $\ket{s}$ and $\ket{\lambda_1 \lambda_2 \ldots \lambda_n \ldots \lambda_N}$ is abbreviated by $\ket{\psi_{n,N}}$, which determines the node reached after $n$ steps. 

To explain what these different qubits encode, recall that at each step the spin can either flip or not flip meaning we can go form one tree to the other or we can stay on the same tree, and the path can either go right or left. At the end of the evolution, if we measure $\ket{\lambda_i}$ in the $\ket{0}$ state it denotes that the path went right at node $i$, while if we measure it in the $\ket{1}$ state, it denotes the path went left at node $i$. For the ket $\ket{s}$, $\ket{0}$ represents spin down and $\ket{1}$ represents spin up. In other words, these qubits uniquely identify a particular node in the two trees. While we keep track whether the path went right or left at each step and we measure this information at the end of the evolution by measuring all the $\lambda_i$ qubits, we do not keep track of whether the the spin flipped or not in a particular step, which is why we can reach the same node with different spin histories.

\begin{figure}[h!]
\[
\Qcircuit @C=0.8em @R=.7em @!R{
\lstick{\ket{\lambda_{N}}}	& \qw		&   \qw			&\qw& \qw 		&\dots & & \gate{U_\downarrow^n} & \gate{U_\uparrow^n}	& \qw			& \meter	\\
\lstick{ \ldots}		&			&  		&& 			& 			& 	 		&  		&&		& \ldots	\\
\lstick{\ket{\lambda_1}}	& \qw	& \gate{U_\downarrow^1}	& \gate{U_\uparrow^1} &\qw &\dots & &\qw 		& \qw		& \qw 			& \meter 	\\
\lstick{\ket{s}}		& \gate{R}	 &  \ctrlo{-1}		& \ctrl{-1} 	&\qw & \dots&	& \ctrlo{-3} 	& \ctrl{-3}		& \gate{R^\dagger} & \meter 	\\
}
\]
\caption{Complete quantum circuit which implements full tree evolution}
\label{fig:circuit}
\end{figure}
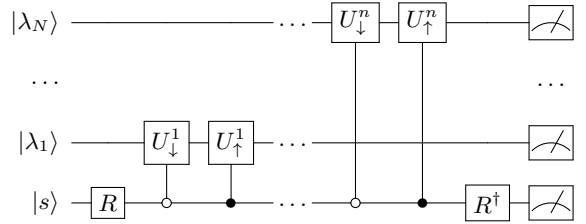

The quantum circuit which implements the evolution is shown in  Fig.~\ref{fig:circuit}. The $\ket{\lambda_i}$ qubits are initialized in the $\ket{0}$ state while the spin qubit, on the other hand, can be initialized in any superposition of $\ket{0}$ and $\ket{1}$. 
The $R$ gate is responsible for rotating into the diagonalized basis, it is given by the $2\times2$ real unitary matrix
\begin{align}
\label{R}
R = \left( \begin{array}{cc}\cos{\lambda} & -\sin{\lambda} \\ \sin{\lambda} &  \cos{\lambda}\end{array} \right)
\,,
\end{align}
while $R^\dagger$ rotates back to the original basis at the end of the evolution before we perform a measurement.
The $U^i_{\uparrow/\downarrow}$ gates are also single qubit operations represented by $2\times 2$ real unitary matrices (we drop the step index for simplicity), which in quantum computing are referred to as $R_Y$ rotations:
\begin{align}
U_\downarrow = \left( \begin{array}{cc}\cos{\theta_\downarrow} & -\sin{\theta_\downarrow} \\ \sin{\theta_\downarrow} &  \cos{\theta_\downarrow}\end{array} \right)  = \left( \begin{array}{cc}\widetilde{A}_{L}^{\downarrow\downarrow} & -\widetilde{A}_{R}^{\downarrow\downarrow} \\ \widetilde{A}_{R}^{\downarrow\downarrow} &  \widetilde{A}_{L}^{\downarrow\downarrow}\end{array} \right) \nonumber \\
U_\uparrow = \left( \begin{array}{cc}\cos{\theta_\uparrow} & -\sin{\theta_\uparrow} \\ \sin{\theta_\uparrow} &  \cos{\theta_\uparrow}\end{array} \right)  = \left( \begin{array}{cc}\widetilde{A}_{L}^{\uparrow\uparrow} & -\widetilde{A}_{R}^{\uparrow\uparrow} \\ \widetilde{A}_{R}^{\uparrow\uparrow} &  \widetilde{A}_{L}^{\uparrow\uparrow}\end{array} \right)
\,,
\end{align}
where we define the basis states on which these matrices act on as  $\ket{0}=\begin{pmatrix}1 \cr 0\end{pmatrix}$ and $\ket{1}=\begin{pmatrix}0 \cr 1\end{pmatrix}$.  

In general the probabilities of the path to go left or right (as well as $\lambda$) could depend on the step, meaning the matrices $U_\downarrow$ and $U_\uparrow$ are different at each step.  If $\lambda$ is different, then $R^\dag(\lambda) R(\lambda')$ operations would need to be inserted between each step. At the end of the circuit evolution, we measure all of the qubits and we record the output. This way we sampled the distribution of final states and generated one event. This corresponds, in our tree notation, to reaching a final tree leaf with definite spin.

\section{Implementation on a Quantum Computer}
\label{sec:implementation}

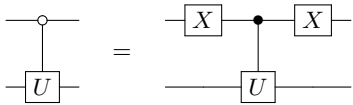
\begin{figure}[h]
\[
\Qcircuit @C=.8em @R=.1em @!R {
& \ctrlo{2} & \qw &  & & \gate{X} & \ctrl{2} & \gate{X} & \qw \\
& & & \push{\rule{.3em}{0em}=\rule{.3em}{0em}} & &&  \\
& \gate{U} & \qw & & & \qw & \gate{U} & \qw & \qw 
}
\]
\caption{Decomposition of a single controlled gate.}
\label{fig:singleControlled}
\end{figure}
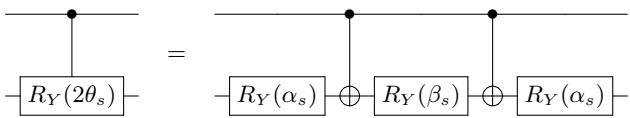
\begin{figure}[h]
\[
\Qcircuit @C=.6em @R=.1em @!R {
& \ctrl{2} & \qw & & & \qw & \ctrl{2} & \qw & \ctrl{2} & \qw & \qw \\
& & & \push{\rule{.3em}{0em}=\rule{.3em}{0em}} & & & & & & \\
& \gate{R_Y(2\theta_s)} & \qw & & & \gate{R_Y(\alpha_s)} & \targ & \gate{R_Y(\beta_s)} & \targ & \gate{R_Y(\alpha_s)} & \qw
}
\]
\caption{The decomposition of the controlled $R_Y$ gate.}
\label{fig:controlledRY}
\end{figure}

In order to run a full quantum simulation of our circuit, and implement it on a currently available test bed, it is necessary to decompose it in terms of single qubit gates and CNOT gates only We have to break down two controlled-$R_Y$ operations, one controlled on $\ket{0}$ and one controlled on $\ket{1}$. We use Fig.~\ref{fig:singleControlled} to relate a unitary transformation controlled on $\ket{0}$ to one controlled on $\ket{1}$,
where $X$ is the standard CNOT gate.  
To break down the controlled-$R_Y$ gate, one uses the decomposition shown in Fig.~\ref{fig:controlledRY}, 
with
\begin{align}
\alpha_s = \frac{\theta_s}{2}\ \ \ \beta_s =- \theta_s
\,.
\end{align}

Combining the above results our quantum circuit can be decomposed in terms of $2+12N$ standard qubit gates (single qubit gates and CNOT gates), showing that the number of gates scales linearly with the number of steps.

\section{Numerical Results}
\label{sec:results}

This section shows some numerical results for simulations of the quantum algorithm and how it compares with a naive classical MCMC implementation.  The quantum circuit is implemented with Qiskit~\cite{qiskit}.  To compute the distributions of various observables, the algorithm is run many times and each measured outcome (leaf and final spin) is recorded.  With these `events', it is possible to then compute the distribution of any observable.   For illustration, two observables are considered: the number of times the system moved left and the first time the system moved left.  For these illustrations, the state always starts as spin down.

A naive classical MCMC is constructed by sampling from the squared amplitudes at each step.  This classical simulation does not contain any interference effects and is therefore expected to produce the incorrect probability distributions for a generic observable when $\lambda\neq 0$.  

\begin{figure}[h!]
\centering
\includegraphics[width=0.23\textwidth]{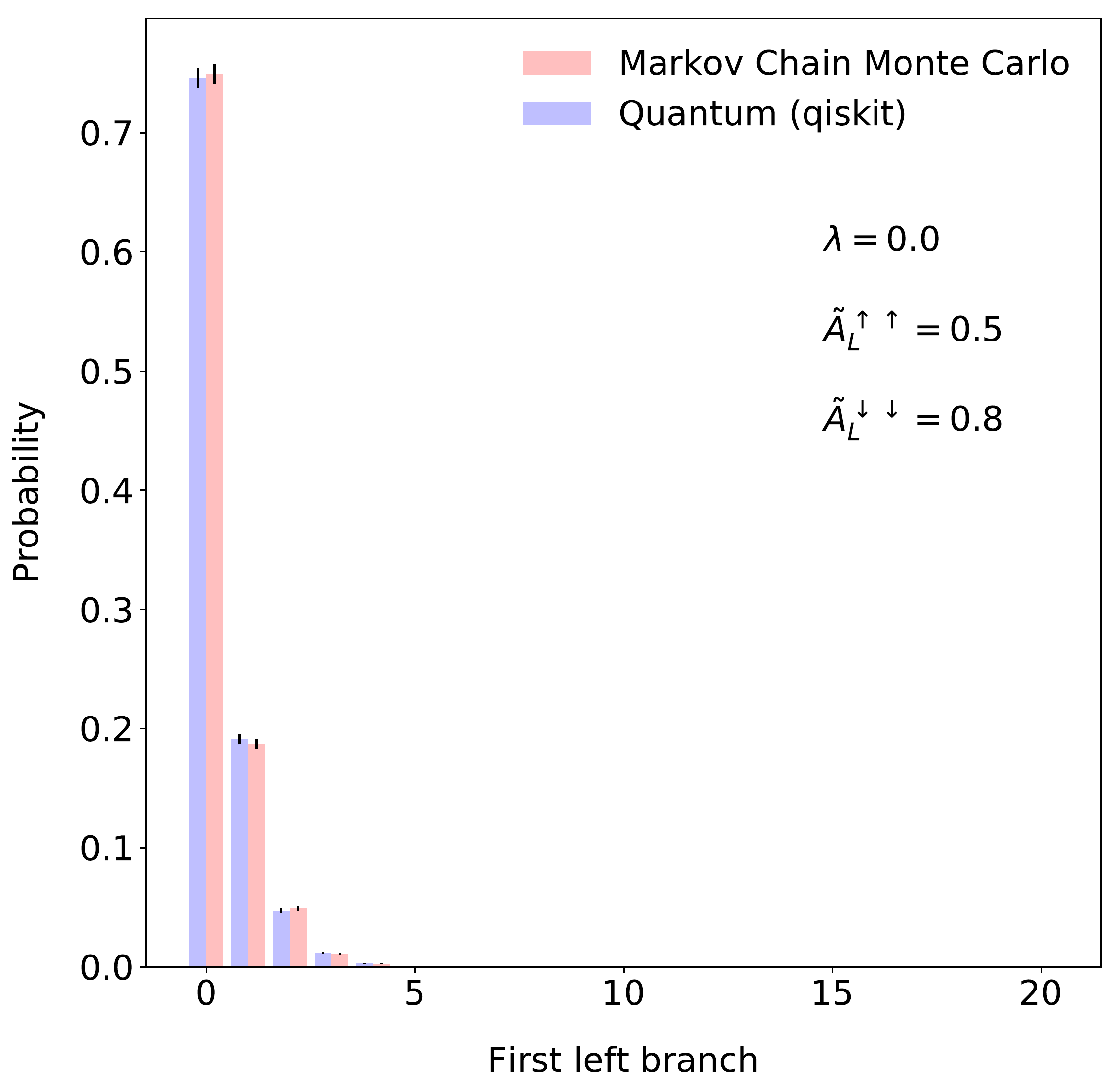}
\includegraphics[width=0.23\textwidth]{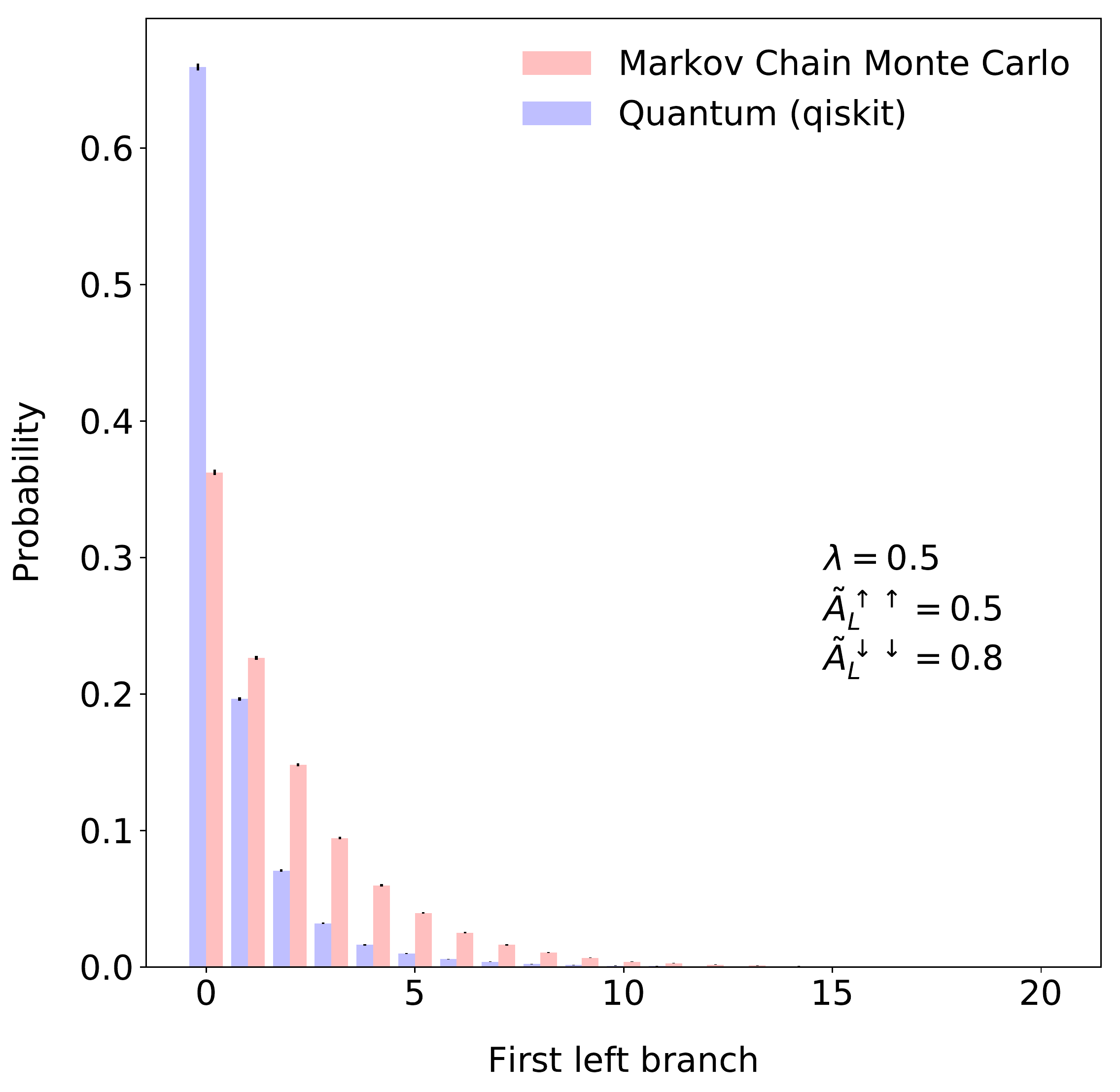}
\includegraphics[width=0.23\textwidth]{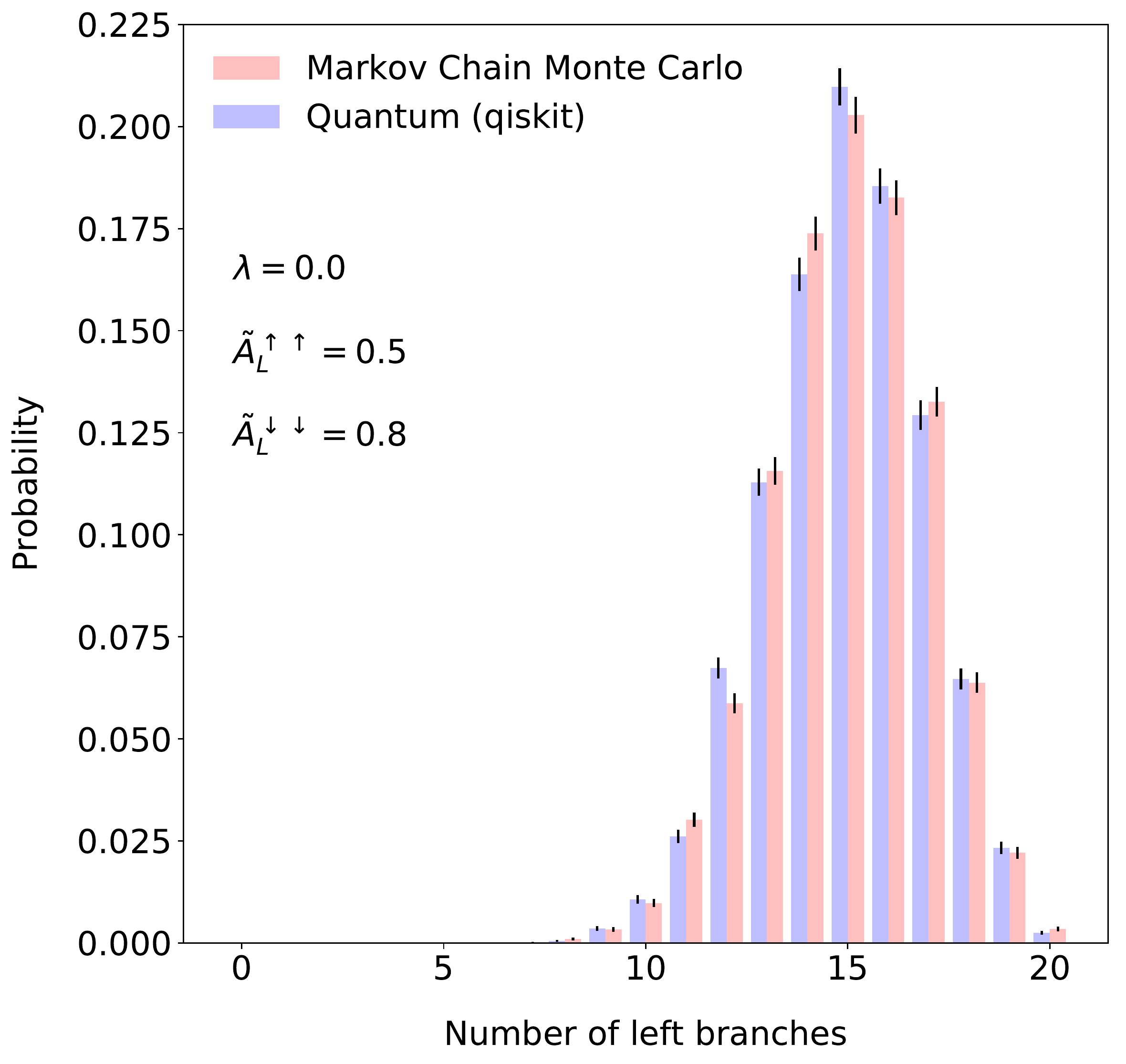}
\includegraphics[width=0.23\textwidth]{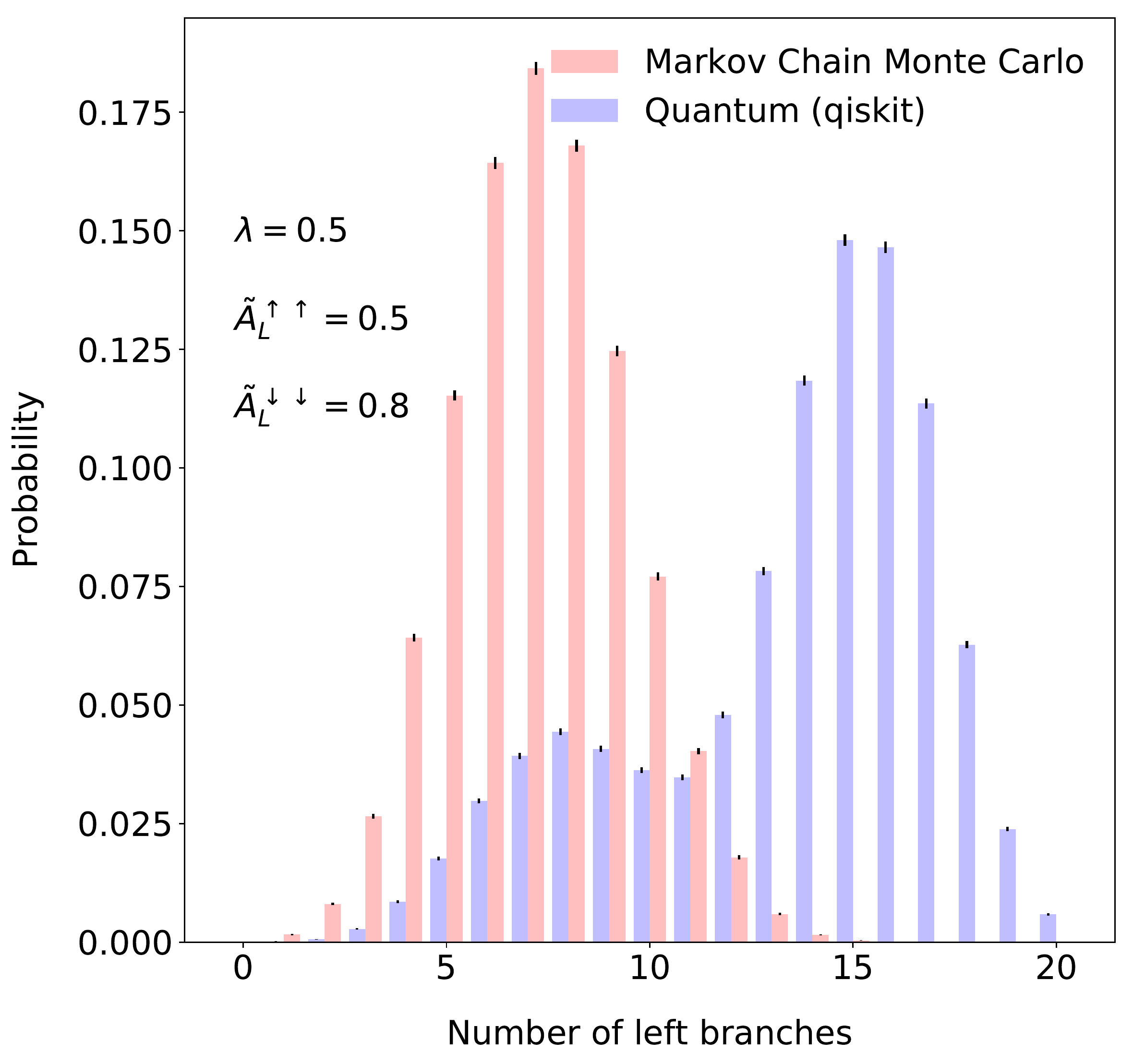}
\caption{Top: histograms of the first depth in which the tree goes left.  Bottom: the number of left branches from the entire tree.  Left: $\lambda=0$ (no interference effects).  Right: $\lambda=0.5$.  Error bars correspond to Poission uncertainties from the finite simulation.  }
\label{fig:hists}
\end{figure}

\begin{figure}[h!]
\includegraphics[width=0.25\textwidth]{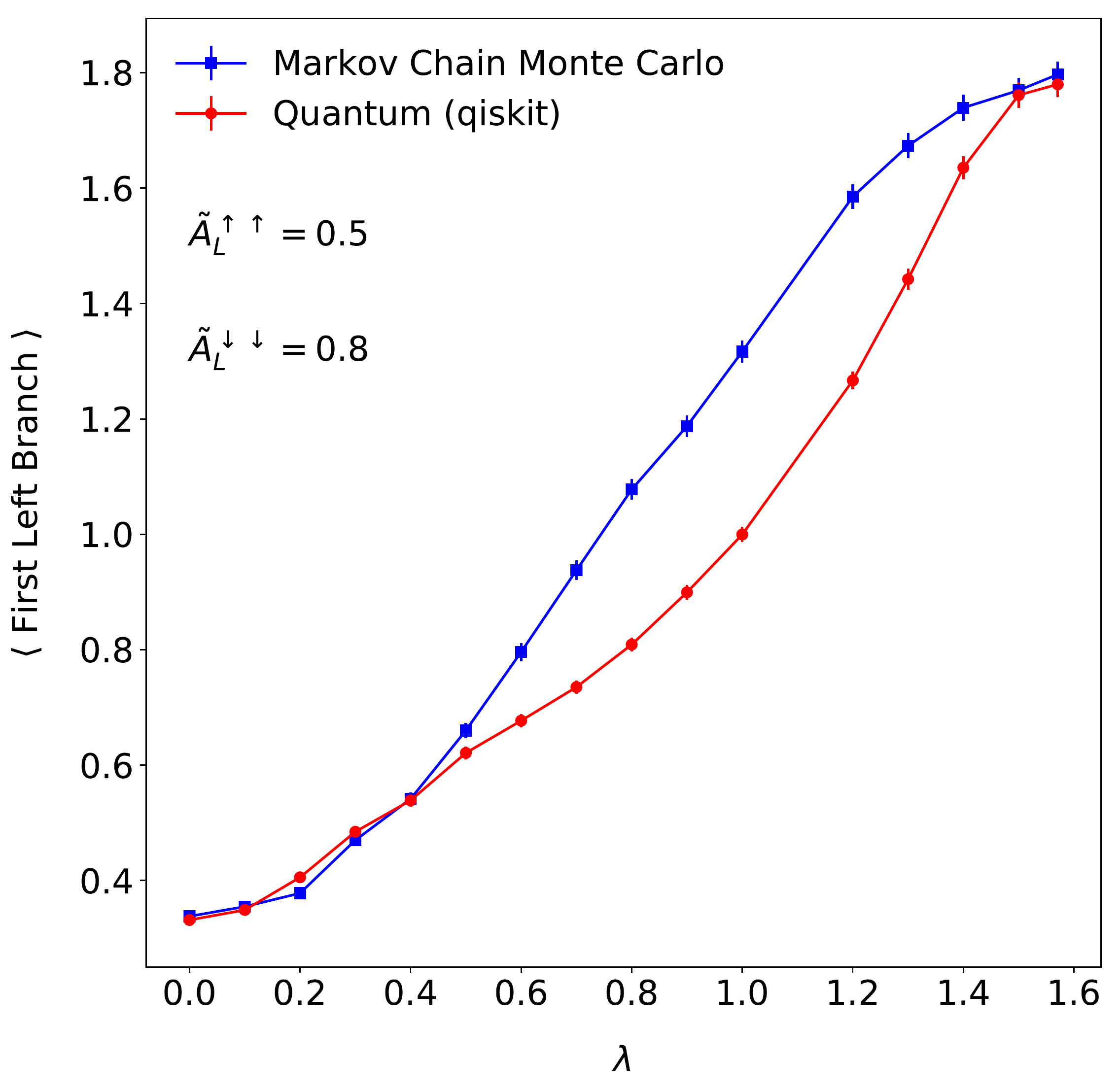}\includegraphics[width=0.25\textwidth]{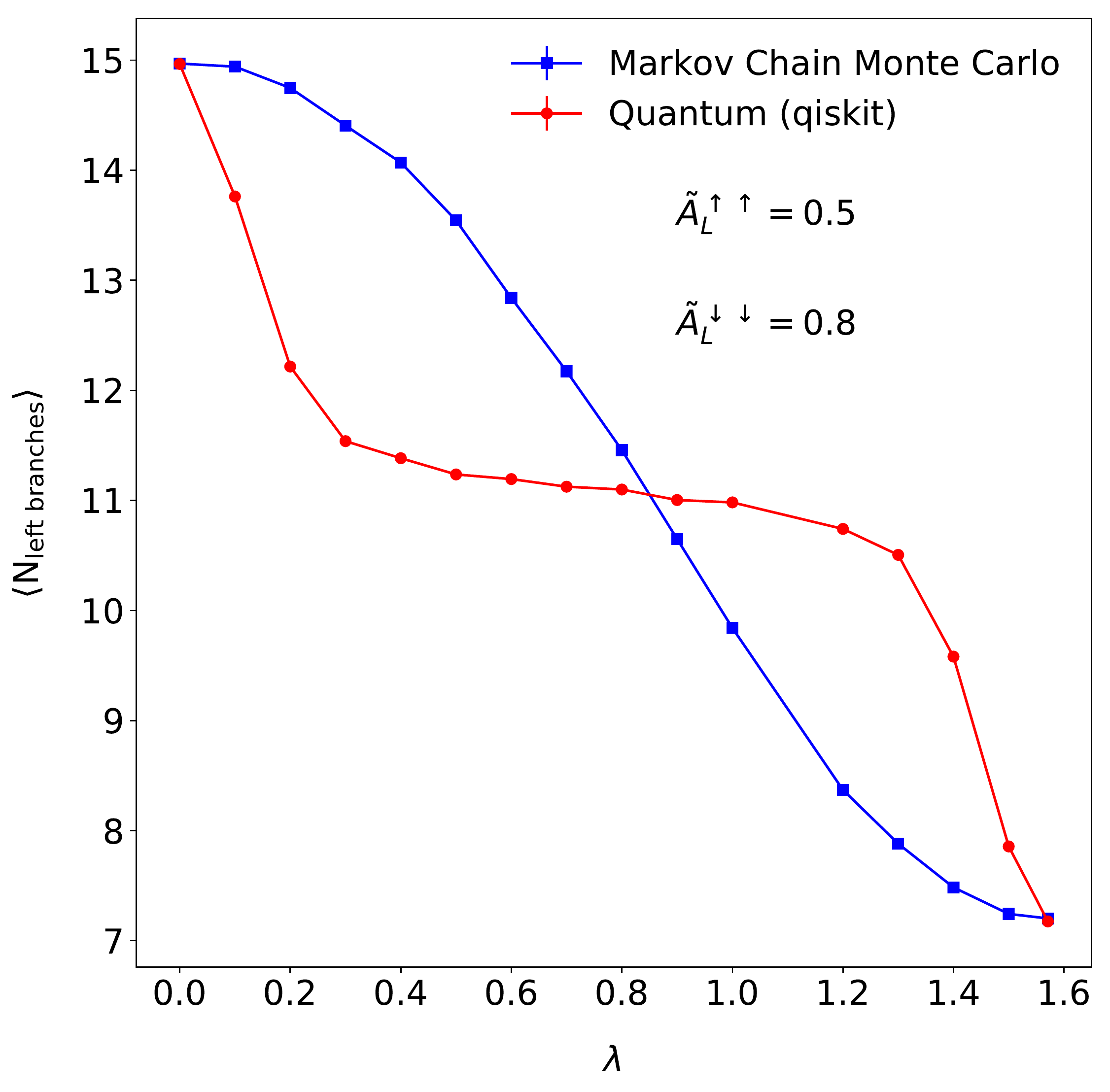}
\caption{Left: the expectation value of the depth of the first left branch as a function of $\lambda$.  Right: the expectation value for the number of total left branches as a function of $\lambda$. Error bars correspond to Poission uncertainties from the finite simulation.   }
\label{fig:nemissions}
\end{figure}

We run our simulations with $N = 20$, with constant values of $\cos^2(\theta_\uparrow)=0.5$ and $\cos^2(\theta_\downarrow)=0.8$.  Figure~\ref{fig:hists} shows histograms of the two observables for different values of $\lambda$ while Fig.~\ref{fig:nemissions} shows how the expectation values for the two observables scales with $\lambda$.


 As expected, the expectation values are the same for the naive MCMC and for the quantum algorithm when $\lambda=0$, but differ as interference effects are introduced.  We have verified that the results from the quantum algorithm agree with the calculation of the full probability distribution using the exponentially scaling method introduced in Sec.~\ref{sec:classical}.  The difference between the MCMC and the quantum algorithm also goes to zero as $\lambda \to \pi/2$, in which case the spin flips at each step in a deterministic way and thus there are no interference effects.

\section{A quantum-inspired classical algorithm}
\label{sec:quantumclassical}

\begin{figure}[h]
\[
\Qcircuit @C=0.4em @R=.5em @!R{
\lstick{\ket{\psi}}	&	\qw& \multigate{1} {\rm U_1}	& \qw & \measureD{\ket{0}} &   &  &    \multigate{1} {\rm U_2}& \qw 	& \measureD{\ket{0}} &  &  &	\ldots & & & \multigate{1} {\rm U_N}	& \qw&\qw	& \meter \\
\lstick{\ket{f}}	& \gate{R}	& \ghost{\rm U_1}	   & 	\qw  &  \qw & 	\qw  & \qw & \ghost{\rm U_2}	 & 	\qw  &  \qw   & 	\qw  & & \ldots & & &  \ghost{\rm U_N}	   & 	\qw  &\gate{R^\dagger}	&    \meter
}
\]
\caption{A quantum algorithm that implements the circuit shown in Figure~\ref{fig:circuit}; each $U_i$ corresponds to the two controlled operations in the earlier circuit.}
\label{fig:2qubit}
\end{figure}
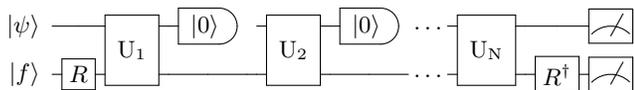

At each step the $U_\downarrow$ and $U_\uparrow$ gates are conditionally applied to a new qubit, but after that the qubit is left alone until the final measurement at the end of the evolution. Therefore, at each step one could measure the qubit on which the  $U_{\downarrow/\uparrow}$ gates act on, store the result in a classical register, reset it to the initial $\ket{0}$ state and reuse it for the next step. Using this method of repeated measurements and resetting the measured qubits one can rewrite the circuit in terms of just two qubits as shown in Figure~\ref{fig:2qubit}.
At each step one records the measurement on the second qubit, and at the very end the first qubit is measured. The combination of these measurements makes up one event. Note that because this circuit can be implemented using just 2 qubits, one can in fact find an efficient quantum inspired classical algorithm. 

We show how this classical algorithm works by considering the $k^\text{th}$ step. Because the second qubit is reset to $\ket{0}$ after each step, the state at the beginning of the step (before $U_k$ is applied) will always be of the form
\begin{align}
\ket{\psi_k}= \begin{pmatrix} a_{1}^{(k)} \\ 0 \\ a_{3}^{(k)} \\ 0 \end{pmatrix}.
\end{align}
After applying the $U_k$ operation through matrix multiplication one finds
\begin{align}
\ket{\psi_k} \rightarrow U_k \ket{\psi_k} = \begin{pmatrix} b_{1}^{(k)} \\ b_{2}^{(k)}  \\ b_{3}^{(k)}  \\ b_{4}^{(k)}  \end{pmatrix},
\end{align}
where the $b_{i}^{(k)}$ are determined from the $a_{1}^{(k)}$ and $a_{3}^{(k)}$ through multiplication with the matrix $U_k$. 

From this one finds that the probabilities $P_0$ and $P_1$ to measure the second qubit as $\ket{0}$ or $\ket{1}$are given by
\begin{align}
P_0 = {b_{1}^{(k)}}^2 + {b_{3}^{(k)}}^2
\,, \qquad 
P_1 = {b_{2}^{(k)}}^2 + {b_{4}^{(k)}}^2
\end{align}
The corresponding states after resetting the second qubit to $\ket{0}$ are given by
\begin{align}
\ket{\psi_k}_0 =  \frac{1}{\sqrt{P_0}} \begin{pmatrix} b_{1}^{(k)} \\ 0 \\ b_{3}^{(k)}  \\ 0  \end{pmatrix}
\,, \qquad 
\ket{\psi_k}_1 =  \frac{1}{\sqrt{P_1}} \begin{pmatrix}  b_{2}^{(k)} \\ 0 \\ b_{4}^{(k)} \\ 0  \end{pmatrix}.
\end{align}

Both of these states have form
\begin{align}
\ket{\psi_{k+1}}= \begin{pmatrix} a_{1}^{(k+1)} \\ 0 \\ a_{3}^{(k+1)} \\ 0 \end{pmatrix},
\end{align}
which has exactly the same form of the state we started with, so that this process can be repeated again.

The same result can be obtained classically by the following classical algorithm for generating a single event, which starts from a fermion in the superposition $f = a f_1 + \sqrt{1-a^2} f_2$, and where we have defined the matrix $U \equiv U_k$. The event is stored in the classical register $c_f$ holding the type of fermion and $c_\psi[\text{step}]$, which holds whether a left branch happened at the given step.   The procedure is described algorithmically in Alg.~(\ref{algo:qica}).

\begin{algorithm}
\begin{algorithmic}
	\STATE{Create empty vector for classical register $c_\psi[m]$}
	\STATE{Set $a_{1} = a$ and  $a_{3} =   \sqrt{1-a^2}$}
	\FOR{$\text{step} = 1 \ldots m$}
	    \STATE{Set $b_{i} = U_{ij} a_{j} $}
	    \STATE{Set $P_0 = (b_{1}^2 + b_{3}^2)$ and $P_1 = b_{2}^2 + b_{4}^2$}
	    \IF{${\rm rand}() < P_0 $}
	    	\STATE{$c\text{[step]} = 0$}
		\STATE{$a_1 = b_1 / \sqrt{P_0}$ and $a_3 = b_3 / \sqrt{P_0}$ }
	    \ELSE
	    	\STATE{$c[\text{step}] = 1$}
		\STATE{$a_1 = b_2 / \sqrt{P_1}$ and $a_3 = b_4 / \sqrt{P_1}$ }
	    \ENDIF
	\ENDFOR
	\IF{${\rm rand}() < a_1^2 / (a_1^2 + a_3^2)$}
	    \STATE{$c_f = 0$}
	\ELSE
	    \STATE{$c_f = 1$}
	\ENDIF
\caption{Quantum inspired classical algorithm.}
\label{algo:qica}
\end{algorithmic}
\end{algorithm}

\section{Conclusions}
\label{sec:conclusions}

In this work, we have introduced a system similar to the quantum walk which smoothly interpolates between a binary tree, amenable to naive classical MCMC approaches, and interfering trees with non-trivial quantum phenomenology.   When non-trivial interference effects are introduced, a classical calculation of all possible outcomes scales exponentially with the depth of the tree.  We have introduced a quantum algorithm that uses an innovative remeasuring technique to sample from the interfering trees with polynomial scaling with the depth of the tree.  In addition to constructing an explicit quantum circuit to implement the algorithm, some numerical results were presented with a simulated quantum computer.  Interestingly, the simple nature of the quantum algorithm inspired a classical approach that is still a Markov chain and thus efficient. 

Given the wide-ranging applicability of classical random walks and quantum walks to aiding complex algorithms, it is likely that the algorithm presented here will be a useful addition to the quantum toolkit.  The application of the interfering trees algorithm and its variations to empowering MCMC algorithms of physical systems could empower many body simulations where quantum effects were previously ignored.  More complex simulations and calculations will also be possible as quantum software and hardware continue to improve.

\acknowledgments
This work is supported by the DOE under contract DE-AC02-05CH11231. In particular, support comes from Quantum Information Science Enabled Discovery (QuantISED) for High Energy Physics (KA2401032).  

\bibliography{myrefs}

\end{document}